\documentclass[final,1p,times]{elsarticle} 
\usepackage{graphicx} 
\usepackage{amssymb} 
\usepackage{amsthm} 
\usepackage{lineno} 

\journal{Nuclear Physics A} 
\begin{document} 

\begin{frontmatter} 


\title{Experimental Studies of Hadronization and Parton Propagation in the Space-Time Domain}

\author{W. K. Brooks and H. Hakobyan}

\address{Departamento de F\'isica, Universidad T\'ecnica Federico Santa Mar\'ia, \\1680 Avda. Espa\~na, Casilla 110-V, Valpara\'iso, Chile}

\begin{abstract} 

Over the past decade, new data have become available from DESY, Jefferson Lab, Fermilab, and RHIC that connect to parton propagation and hadron formation. Semi-inclusive DIS on nuclei, the Drell-Yan reaction, and heavy-ion collisions all bring different kinds of information on parton propagation within a medium, while the most direct information on hadron formation comes from the DIS data. Over the next decade one can hope to begin to understand these data within a unified picture. We briefly survey the most relevant data and the common elements of the physics picture, then highlight the new Jefferson Lab data from CLAS, and close with prospects for the future.

\end{abstract} 

\end{frontmatter} 



Studies of QCD in the space-time domain have received an increasing amount of attention over the past decade due to the availability of several new, precise datasets. Use of atomic nuclei in high energy collisions brings the element of a QCD system that is extended in size, so that the time development of the resulting interactions has observable consequences. In the case of cold nuclei, the time development of fundamental processes can be studied in detail using the well-understood nuclear system as a spatial analyzer. In the case of relativistic heavy ion collisions, the same processes can be used as tools to explore the nature of the hot dense matter that is formed. It is to be hoped that the studies of cold QCD matter, once matured, can influence the interpretation of what is seen in the hot dense systems, in addition to their intrinsic interest for QCD.

This article discusses present and future studies of semi-inclusive deep inelastic scattering (DIS) from DESY and from Jefferson Lab, the Drell-Yan (D-Y) reaction in proton-nucleus collisions from Fermilab, and the potential connections to heavy-ion data from RHIC and LHC.

\begin{figure}[ht]
\centering
\includegraphics[height=5.5cm]{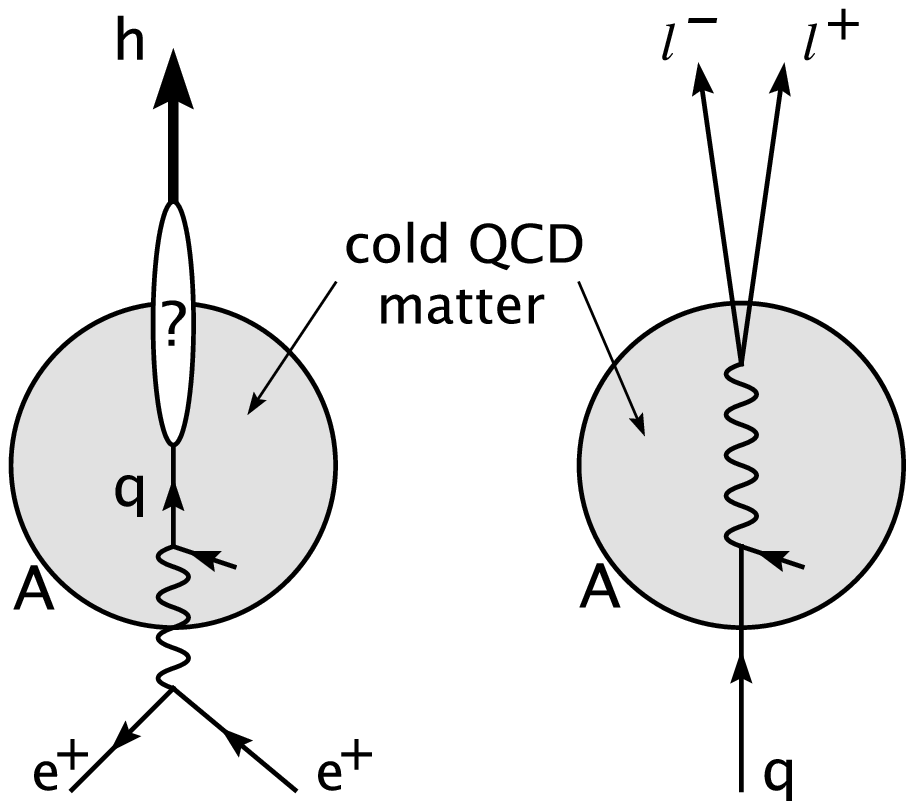}
\hspace*{.3cm}
\includegraphics[height=5.5cm]{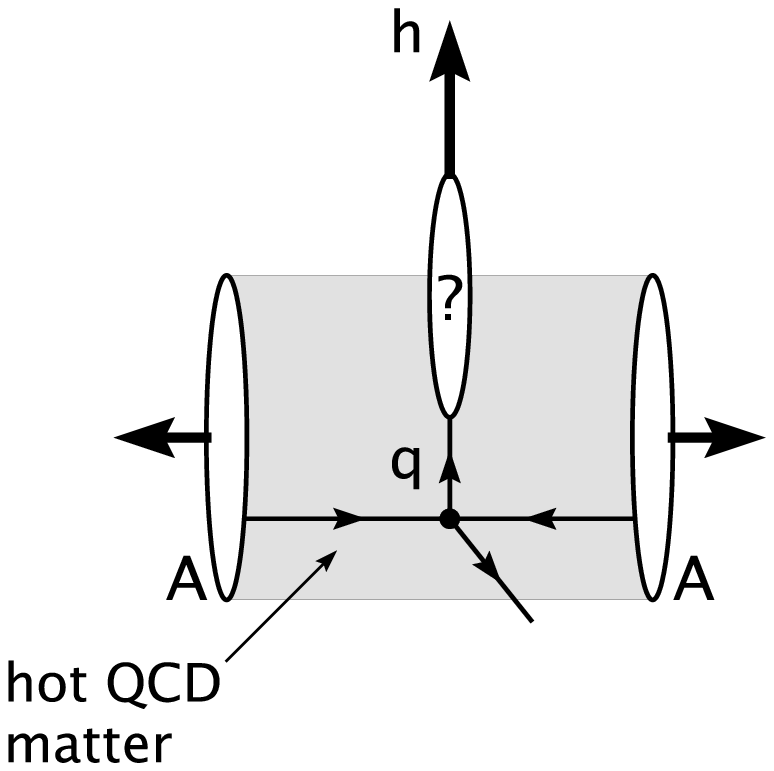}
\caption[]{Quark propagation inside a target nucleus (``cold QCD matter'')
   in lepton-nucleus ({\it left}) and the Drell-Yan process ({\it center}) in hadron-nucleus collisions. {\it Right:} Hard scattered parton traveling through the
   ``hot QCD matter'' produced in a nucleus-nucleus collision. Taken from Ref. \cite{Accardi09a}.
 \label{fig1}
}
\end{figure}

\section{Physical Pictures}\label{physicalpictures}

The conceptual pictures of the processes discussed here are illustrated in Fig.~\ref{fig1}, which is taken from a recent comprehensive review~\cite{Accardi09a}. In the DIS process on the left in this figure, a virtual photon from an incident lepton is absorbed by a quark within a nucleus; the highly virtual colored quark propagates over some distance through the nuclear medium, evolves into a color-neutral 'prehadron' and subsequently emerges as an hadron. Strictly speaking, for this picture to be valid, Bjorken $x$ must be greater than 0.1 to avoid quark pair production\cite{Brodsky92}; further, from a naive classical perspective, the prehadron and hadron may be formed inside the nucleus or outside the nucleus. The center diagram depicts the D-Y process in which a quark from an incident hadron annihilates with an antiquark from the nucleus, forming a photon which subsequently materializes as a dilepton pair. This pair carries information about the incident quark's passage through the nuclear medium. The diagram on the right illustrates a heavy-ion collision in which a scattered parton from the collision propagates over a distance within the hot dense medium, then evolves into a prehadron and a hadron at later times. At a superficial level, all three pictures contain a propagating quark within a strongly interacting medium, and two contain the process of a hadron forming, thus, there may be strong connections between the three that could be explored. In more detail, there are some differences: initial state and final state interactions enter differently, and several characteristics of the hot and cold media are different. Nonetheless, the fundamental process by which the quark interacts with the medium is gluon exchange in all cases, the formation of the hadron must proceed through a color singlet state that ultimately fulfills the requirements of confinement, and each step has its own characteristic time scale. The next steps in understanding these processes in cold matter include determination of characteristic time scales for the quasi-free quark lifetime and for the hadron formation, as well as parameters such as the transport coefficient $\hat{q}$ that characterize the quark's interaction with the medium. These goals require a deeper understanding of the mechanisms of hadron formation, including the roles of coherence and quantum interference.

\begin{figure}[ht]
\centering
\includegraphics[height=8cm]{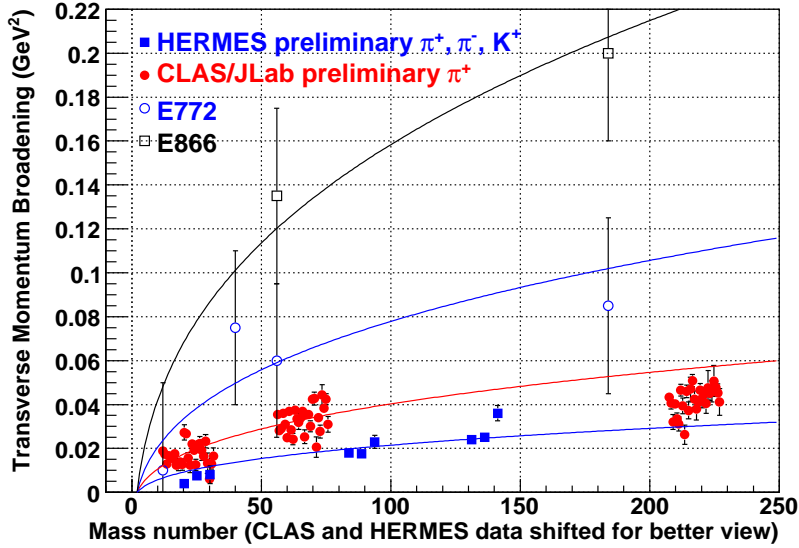}
\caption[]{Transverse momentum broadening measurements from the Drell-Yan process in proton-nucleus scattering and from semi-inclusive DIS measurements. The target masses shown are 12, 40, 56, 184 (D-Y); 20, 84, 131 (HERMES); and 12, 56, 207 (CLAS). 
 \label{fig2}
}
\end{figure}

\section{Transverse Momentum Broadening}

An observable that is common to the DIS and D-Y processes is {\em{transverse momentum broadening}}. It is defined as the increase in the square of the momentum component transverse to the nominal initial direction of the quark after it has passed through a strongly interacting medium. This observable is expected to be a good measure of partonic multiple scattering and potentially its associated medium-stimulated gluon bremsstrahlung. This can be discussed in terms of the {\em{quark}} transverse momentum change $\Delta k_T^2$ or as that of the observed {\em{final state hadron}}, such as a pion, $\Delta p_T^2$. To a very good approximation these are related by 
\begin{equation}
\Delta p_T^2 = z_h^2 \Delta k_T^2
\end{equation}

\noindent as noted in \cite{Domdey08}. In this paper this will be referred to as {\em{z-scaling}}.

In Fig.~\ref{fig2} is plotted the transverse momentum broadening seen in recent DIS and D-Y experiments on nuclei. The plot shows data from the E772 and E866 D-Y experiments at Fermilab, the preliminary HERMES data \cite{HERMESpT} from DESY, and the preliminary CLAS data \cite{CLASpT,E02104} from Jefferson Lab. The data from HERMES and CLAS have been shifted toward the right of the correct mass number so that the small error bars can be seen. Each of the 81 CLAS points shown is in a bin in $\nu$, $Q^2$, and $z_h$ for $\pi^+$ for carbon, iron and lead targets, while the HERMES points  are for $\pi+$, $\pi^-$, and $K^+$ for neon, krypton, and xenon targets. The four solid lines correspond to a function proportional to the mass number to the $1/3$ power, that is, proportional to the average nuclear diameter. While the precision of the D-Y data is somewhat limited, the high precision of the new DIS data is clearly impressive. The difference between the CLAS data and the HERMES data can be explained by the z-scaling mentioned above, since the two data sets are measured at different average values of $z_h$.

\begin{figure}[ht]
\centering
\includegraphics[height=6.5cm]{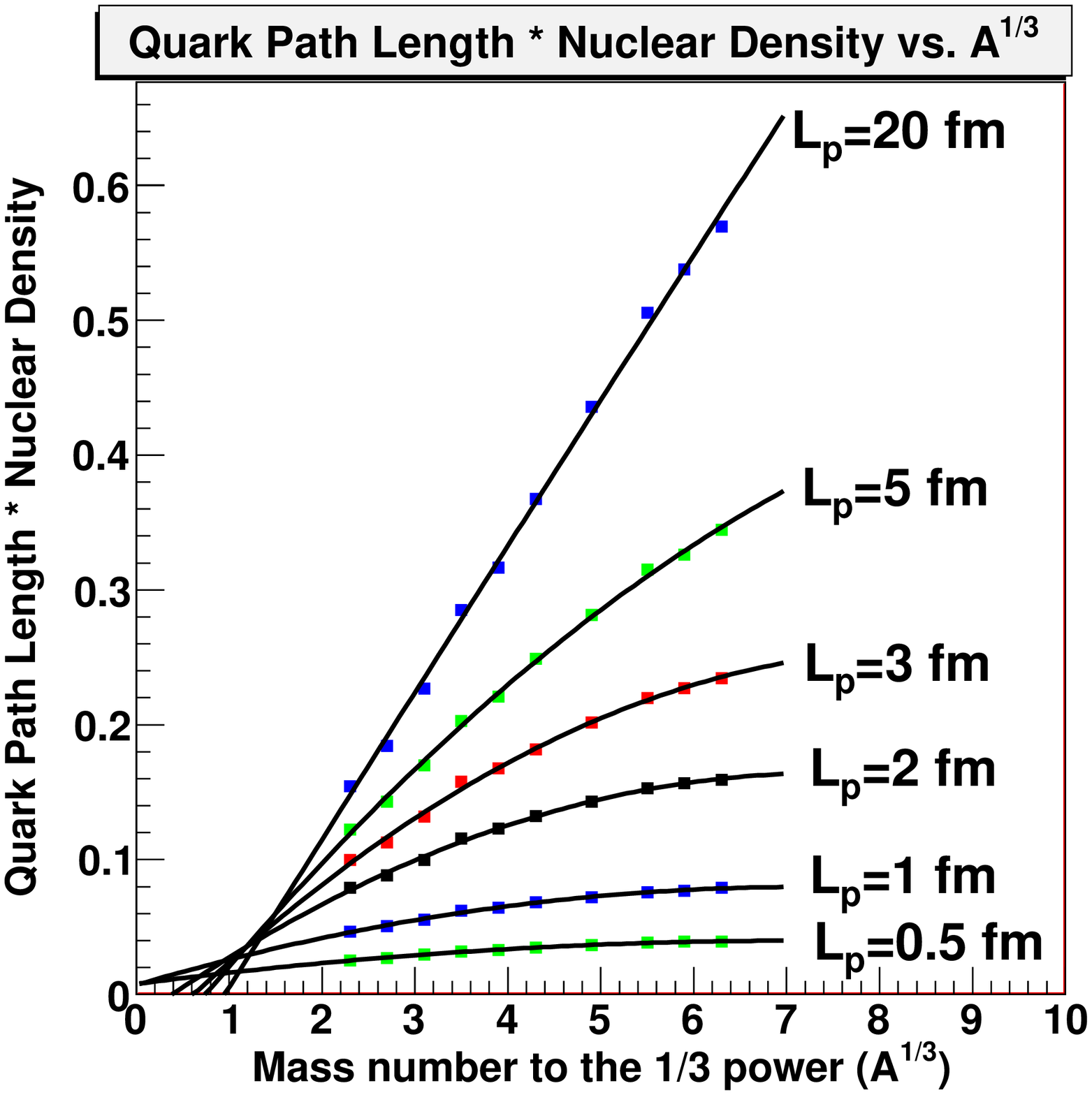}
\includegraphics[height=6.5cm]{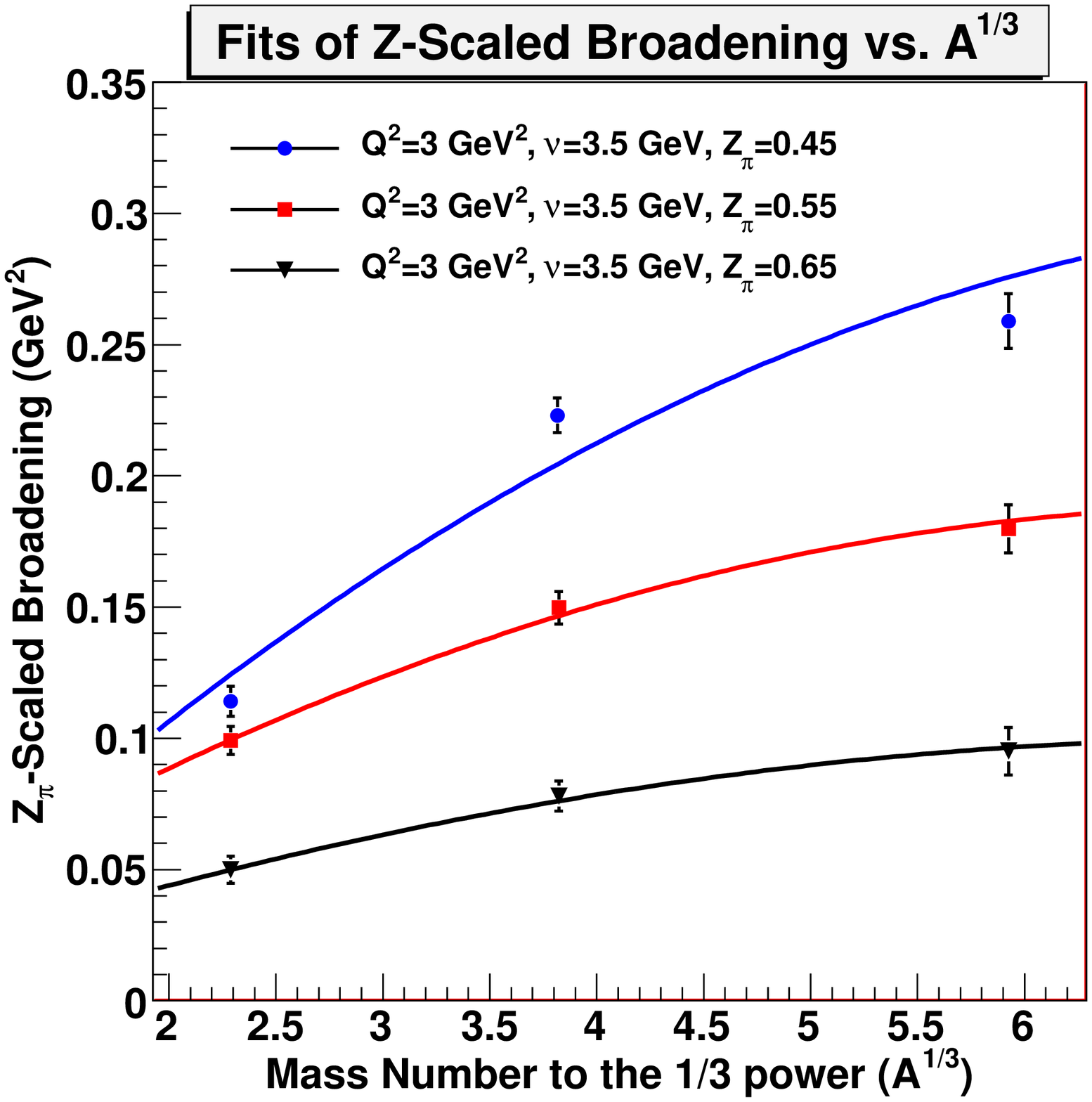}

\caption[]{Left: effect of nuclear geometry in producing non-linear broadening curves. This effect, which only exists at lower energies, provides measurement sensitivity for determination of production lengths. Right: fits of a geometric Monte Carlo model to the CLAS/JLab data for $\pi^+$ transverse momentum broadening in three-dimensional kinematic bins in $Q^2$, $\nu$, and $z_h$. The error bars shown only reflect statistical uncertainties. 
 \label{fig3}
}
\end{figure}

One feature of the CLAS data is especially noteworthy. The points for the heaviest target fall systematically below the line, indicating a saturation behavior. This would be expected in the case that the quasi-free quark evolves into a prehadron within the medium for the largest nucleus; since transverse momentum broadening ceases at the point of color neutralization, the broadening is no longer proportional to the nuclear diameter in the larger nuclei. If this picture is correct, then at these lower energies there is experimental sensitivity to measure the effective lifetime of the quasi-free quark, usually referred to as the {\em{production time}} $\tau_p$. At higher energies, the broadening is expected to be simply proportional to the medium thickness, and this sensitivity is lost. This effect is illustrated in the left panel of Fig.~\ref{fig3}, which shows the effect of geometry on data of this kind using a geometric Monte Carlo model with realistic nuclear density distributions. In the case when the production length $\tau_p$ is longer than the dimensions of any of the nuclei, there is a linear relationship between the average broadening and the nuclear radius. If, however, $\tau_p$ becomes less than the diameters of the nuclei, it introduces a reduction in the average broadening and the straight line acquires a curvature. This shape can in principle be used to measure the production length. 

As an illustration of this idea, the shapes from the geometric Monte Carlo have been fitted to the CLAS data for pT broadening which have been scaled by the $z_h$ factor shown in Equation 1. The fit can be seen in the right panel of Fig.~\ref{fig3}; in this plot, the errors are only statistical, and 9 of the 81 CLAS data points are displayed. While it would be premature to make conclusions at this stage, the results are encouraging and suggestive that a procedure of this kind can be used to extract $\tau_p$ as a function of $Q^2$, $\nu$, and $z_h$ from these data. Further, using the z-scaling, one can scale both the HERMES data and the CLAS data to estimate the transport coefficient $\hat{q}$. It can be seen on inspection of Figures \ref{fig2} and \ref{fig3} that the value of $\hat{q}$ thus obtained is quite small, and similar in magnitude to that of Ref. \cite{BDMPSa}. These results can also be analyzed within the color dipole formalism \cite{Dolejsi93} to account for the energy dependence between the D-Y data and the DIS data \cite{Johnson01}. 

The propagation of quarks through a hot dense medium is associated with the observation of jet quenching as well as other related phenomena. Theoretical approaches specific to describing jet modification in hot dense matter have recently been reviewed \cite{Majumder07, Bass09, Wiedemann09}. While the available sophisticated approaches can describe the data, they do so with very different equivalent parameters within the same basic physical picture, and even alternative ideas about the underlying physics can provide an adequate description \cite{Kopeliovich08a}. It is very desirable to have a unique explanation of the experimental data, however, it seems that more experimental constraints may be needed to clarify the situation. These could be provided by the measurements described herein from cold QCD matter to the the extent they can be addressed by similar theoretical techniques, as well as by additional measurements at RHIC; heavy ion collisions at LHC may provide the most stringent tests of the existing approaches.

\begin{figure}[ht]
\centering
\includegraphics[height=6.cm]{majumder.eps}
\includegraphics[height=6.5cm]{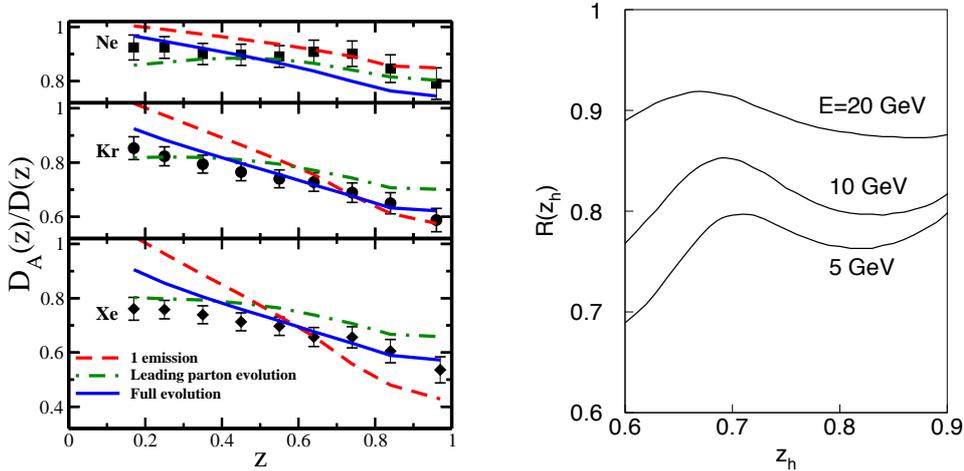}
\caption[]{Theoretical calculations for the hadronic multiplicity ratio. On the left, comparison of HERMES data to a calculation \cite{Majumder09} based on modification of the in-medium fragmentation functions. On the right, a calculation \cite{Kopeliovich08a} of the multiplicity ratio emphasizing the role of quantum interferences in prehadron formation. 
 \label{fig4}
}
\end{figure}

\begin{figure}[ht]
\centering
\includegraphics[height=7.5cm]{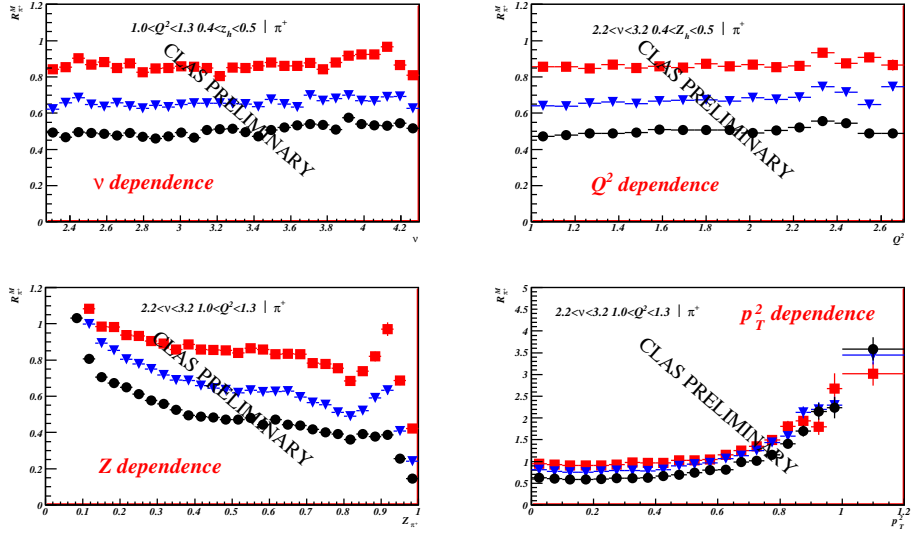}
\caption[]{Preliminary CLAS data for the hadronic multiplicity ratio for positive pions. Each data point is binned in three variables, permitting a multidimensional analysis of this observable.
 \label{fig5}
}
\end{figure}

\begin{figure}[ht]
\centering
\includegraphics[height=7.cm]{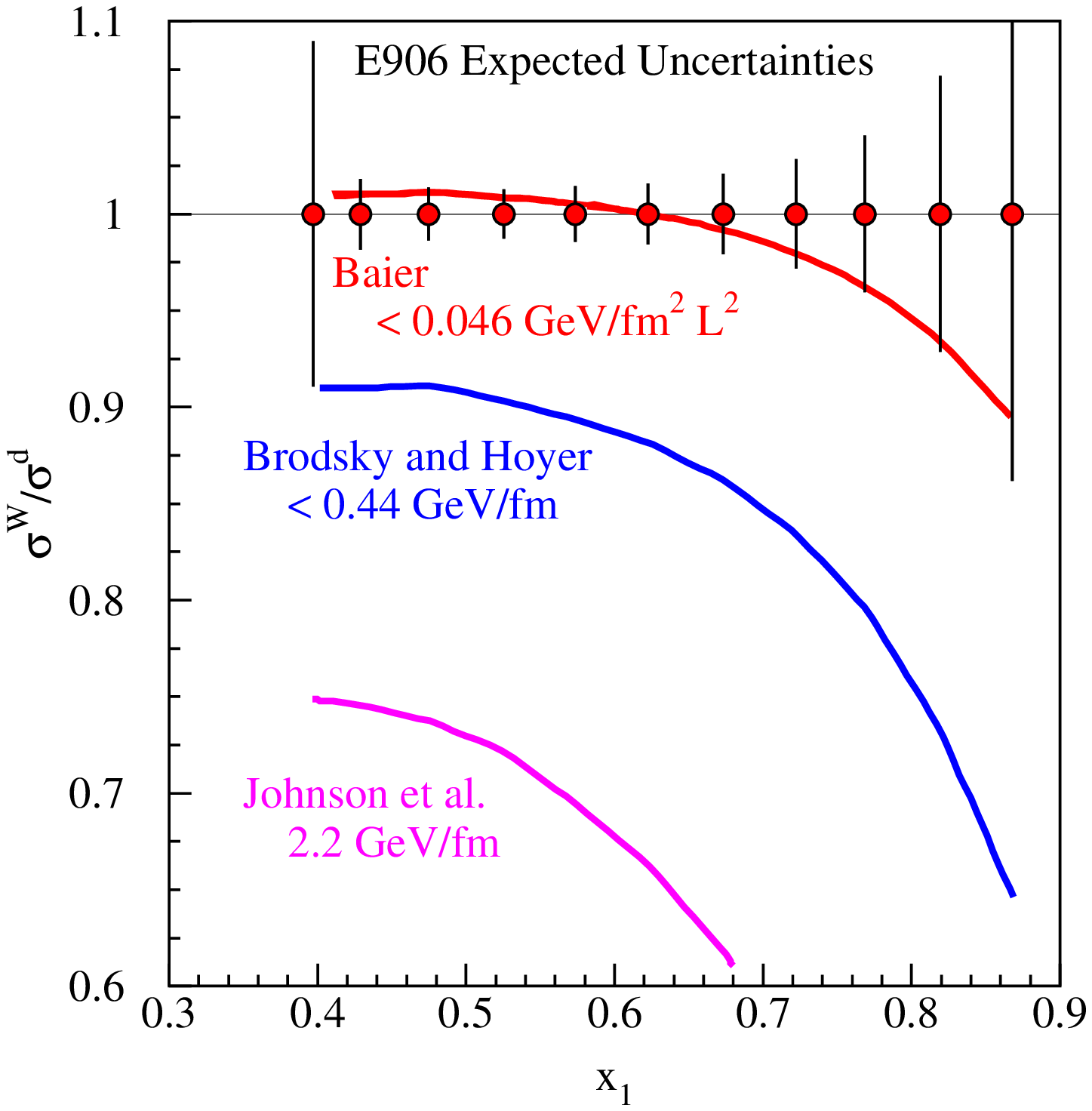}
\caption[]{Anticipated uncertainties in Fermilab experiment E906 which will provide high precision D-Y measurements. The calculations shown are from \cite{BDMPSa, Brodsky93, Johnson02}.
 \label{fig6}
}
\end{figure}

\section{Hadron Attenuation}

The primary access to space-time information on the formation of hadrons comes from semi-inclusive DIS on nuclei. The observable most relevant here is the {\em{hadronic multiplicity ratio}} $R^h_M$, which expresses the normalized change in production yield of hadrons in larger nuclei compared to deuterium or hydrogen. 
The ratio $R^h_M$ can be analyzed within a model to determine the formation lengths of hadrons as a function of the relevant variables, typically $Q^2$, $\nu$, $p_T$, $z$, and $\phi$ (the azimuthal angle of the hadron with respect to the virtual photon direction). 

The hadron attenuation data have been explained in terms of two different processes: modification of the in-medium fragmentation functions \cite{Majumder09, Arleo03} (see Fig.~\ref{fig4} left), and interaction of the color singlet prehadron with the nuclear medium \cite{Kopeliovich04, Falter04, Accardi05, Akopov05, Gallmeister07}. It has thus far not been possible to determine the relative magnitudes of these two processes from the HERMES data \cite{HERMES1, HERMES2}. The CLAS data (see Fig.~\ref{fig5}), with two orders of magnitude more integrated luminosity, a wider range of target masses, and a lower energy, may be able to uniquely determine the relative contributions of these two mechanisms, however, much theoretical work is required in order to hope for a definitive result.

The most recent theoretical development has been a fully quantum mechanical calculation of the multiplicity ratio \cite{Kopeliovich08a} using an improved fragmentation model \cite{Kopeliovich08b}, as shown in Fig.~\ref{fig4} on the right. In this calculation it is found that there can be interferences between the processes of forming the hadron inside and outside the medium. In this exploratory work, this interference generates a change in the shape and magnitude of the functional form of $R_M^h$ which has an energy dependence as can be seen in the figure. 
If confirmed by further theoretical investigations, this may prove to be an important new factor in the determination of this observable. 

\section{Future Prospects}

Several new experimental developments are planned over the next decade that will have an important impact on these studies. Continued progress in understanding the accumulating RHIC data will be ongoing. The full analysis of the existing CLAS dataset for 4-5 hadrons will be completed. The LHC heavy ion running is planned to begin within two years, with a significant extension in collision energy. A new D-Y measurement \cite{DY2010} is planned at Fermilab within the same time frame, which will allow much higher precision data to be obtained, comparable to the DIS measurements. A plot is shown in Fig.~\ref{fig6} that indicates the expected uncertainties for a tungsten target compared to deuterium in this new D-Y measurement.

A major improvement in the DIS data will occur following the completion of the Jefferson Lab upgrade. This upgrade, which is planned for completion in 2015, will provide 11 GeV electron beams to three experimental halls and 12 GeV electron beam for a new fourth hall, Hall D. Large acceptance spectrometers will be available in Hall B for electron beam experiments and in Hall D for photon beam experiments. While the latter can be used with nuclear targets to explore related topics such as the space-time development of color dipoles, the former can be used for a much expanded program of measurements of the type described above at HERMES and with CLAS. These will provide high luminosity measurements for $\nu$=2-9 $GeV$ and $Q^2$=2-9 $GeV^2$ \cite{E1206117} with the CLAS12 spectrometer that has already begun construction. With the planned complement of particle identification, a large number of hadrons will be accessible, as can be seen in Table~\ref{table:hadron_list}, which lists eleven mesons and eight baryons with $c\tau$ longer than nuclear dimensions. Access to relatively low-rate hadrons will be possible with the planned luminosity of more than $10^{35}~/cm^2s$, three orders of magnitude more than the pioneering measurements at HERMES.

\begin{table}[h!]
\begin{center}
\begin{tabular}{ccccccc} \hline 
hadron & $c\tau$ & mass & flavour  & detection & Production rate \\
       &         &(GeV/c$^2$) & content &  channel &  per 1k DIS events \\
                  \hline 
$\pi^0$ & 25 nm & 0.13 & $u\bar{u}d\bar{d}$ & $\gamma\gamma$ & 1100  \\ 
$\pi^+$ & 7.8 m & 0.14 &   $u\bar{d}$ & direct & 1000  \\ 
$\pi^-$ & 7.8 m & 0.14 &   $d\bar{u}$  & direct & 1000  \\ 
$\eta$ & 0.17 nm & 0.55 & $u\bar{u}d\bar{d}s\bar{s}$&$\gamma\gamma$ & 120  \\ 
$\omega$ & 23 fm & 0.78 &  $u\bar{u}d\bar{d}s\bar{s}$ & $\pi^+\pi^-\pi^0$ & 170  \\ 
$\eta'$ & 0.98 pm & 0.96 &  $u\bar{u}d\bar{d}s\bar{s}$ & $\pi^+\pi^-\eta$ & 27  \\ 
$\phi$ & 44 fm & 1.0 &  $u\bar{u}d\bar{d}s\bar{s}$ & $K^+K^-$ & 0.8  \\ 
$f1$ & 8 fm & 1.3 &  $u\bar{u}d\bar{d}s\bar{s}$ & $\pi\pi\pi\pi$ & -  \\ 
$K^+$ & 3.7 m & 0.49 &  $u\bar{s}$ & direct & 75  \\ 
$K^-$ & 3.7 m & 0.49 &  $\bar{u}s$ & direct & 25  \\ 
$K^0$ & 27 mm & 0.50 &  $d\bar{s}$ & $\pi^+\pi^-$ & 42  \\ \hline
$p$ & stable & 0.94 &  $ud$ & direct & 530 \\ 
$\bar{p}$ & stable & 0.94 &  $\bar{u}\bar{d}$ & direct & 3 \\ 
$\Lambda$ & 79 mm & 1.1 &  $uds$ & $p\pi^-$ & 72 \\ 
$\Lambda(1520)$ & 13 fm & 1.5 &  $uds$ & $p\pi^-$ & - \\ 
$\Sigma^+$ & 24 mm & 1.2 &  $us$ & $p\pi^0$ & 6 \\ 
$\Sigma^0$ & 22 pm & 1.2 &  $uds$ & $\Lambda\gamma$ &  11 &\\ 
$\Xi^0$ & 87 mm & 1.3 &  $us$ & $\Lambda\pi^0$ & 0.6  \\ 
$\Xi^-$ & 49 mm & 1.3 &  $ds$ & $\Lambda\pi^-$ & 0.9  \\ \hline 
\end{tabular}
\end{center}
\caption{\small{Final-state hadrons accessible for formation 
length and transverse momentum broadening studies in CLAS12. The 
rate estimates were obtained from the LEPTO event generator for 11 GeV incident electrons.}}
\label{table:hadron_list}  
\end{table}

\section{Conclusions}

A program of measurements has been described that is yielding new insight into space-time properties of QCD and providing new tools for probing microscopic features of strongly interacting systems. The topics of quark propagation and hadron formation naturally provide a connection between experiments with proton beams, lepton beams, and heavy ion collisions, linking measurements at five international-scale laboratories, and stimulating new theoretical activity. The data obtained over the past decade are now beginning to be understood in detail, and the prospects for interesting new data in the next decade are excellent. Measurements using cold QCD matter are now being used to extract such quantities as the effective lifetime of the quasi-free quark and the formation times for a variety of hadrons, as well as gaining new insight into the mechanisms of hadronic fragmentation. The transport coefficient and other transport parameters are being extracted from these data with unprecedented precision. It is hoped that the studies in cold QCD matter will, when matured, be able to influence the interpretation of observations in hot dense matter, in addition to gaining insight into fundamental space-time properties of QCD.




\end{document}